\documentclass[preprint,11pt,aps,amsmath,amssymb,superscriptaddress]{revtex4}
\usepackage[cp1250]{inputenc}

\begin{document}

\title{Towards Loop Quantization of Plane Gravitational Waves}

\author{Franz Hinterleitner}
\affiliation{Department of Theoretical Physics and Astrophysics,
Faculty of Science of the Masaryk University, Kotl\'{a}\v{r}sk\'{a}
2, 611\,37 Brno, Czech Republic}
\author{Seth  Major}
\affiliation{Department of Physics, Hamilton College, Clinton NY
13323 USA}

\date{June 2011}

\begin{abstract}
The polarized Gowdy model in terms of Ashtekar-Barbero variables is further reduced by including the Killing equations for plane-fronted parallel gravitational waves with parallel rays. The resulting constraint algebra, including one constraint derived from the Killing equations in addition to the standard ones of General Relativity, are shown to form a set of first-class constraints. Using earlier work by Banerjee and Date the constraints are expressed in terms of classical quantities that have an operator equivalent in Loop Quantum Gravity, making space-times with pp-waves accessible to loop quantization techniques. 
\end{abstract}

\maketitle PACS 04.20.Fy, 04.30.Nk, 04.60.Ds, 04.60.Pp

\section{Introduction}

This paper on the formulation of plane-fronted gravitational waves with parallel rays (pp-waves) in the connection formulation has two main motivations. First, space-times with pp-waves are interesting objects to test methods of quantizing gravity. Like midi-superspace models these space-times, being homogenous in two directions and inhomogeneous in the third direction, have a degree of difficulty lying between the complicated full theory of General Relativity (GR) and homogenous cosmological models amenable to quantization. Second, the recent abundance of investigations of conjectured Lorentz invariance violations, including an energy-dependent speed of light, deformations of special relativity, and experimental bounds on such effects highlight the lack of {\em ab initio} calculations of these effects from fundamental theories. Some of these conjectures are inspired by the granularity of space predicted by Loop Quantum Gravity (LQG) \cite{RSareavol,Lvol,ALareavol,Tlength,Mangle,Blength}. However the full theory seems to satisfy local Lorentz covariance \cite{RSlorentz,Wlorentz}. To our knowledge, there are no calculations in tractable models that clearly reconcile the granularity of spatial geometry with local Lorentz symmetry. (Although Bojowald and Hossain \cite{MB1} provide a derivation of gravitational wave dispersion in LQG in a cosmological context using perturbative methods.) Gravitational pp-waves appear to be simple enough for deducing a definitive answer - at least in the context of a model system - to the question of whether we should expect dispersion of these waves in the quantum gravity resulting from loop quantization techniques.

It is well known that attempts of (loop) quantization of simplified models suffer from the problem that quantized reduced models have different features than the full theory - the more a system is simplified before quantization, the more likely that genuine features of full GR will be lost. In this paper we start, as a feasible compromise, from a slightly more general setting than pp-waves: The polarized Gowdy model in a form prepared for loop quantization, thoroughly studied and presented by Banerjee and Date \cite{BD1,BD2}.  Like space-times inhabited by pp-waves, this model is homogenous in two dimensions. But the Gowdy model has a different global topology. It is compact, whereas our model has the global topology of Minkowski space. Nonetheless due to homogeneity, we can choose an arbitrary finite area from the plane wave fronts and, as we consider only finite wave packets, the global topology of the inhomogeneous direction is not relevant for our purpose. 

The essential step in the reduction from the polarized Gowdy model to pp-waves is to single out space-times with waves traveling into one direction and so to avoid colliding plane waves that lead to well-known complicated interaction processes and eventually to singularities \cite{GP}. One such reduction of the polarized Gowdy model to pp-waves is carried out in \cite{we}. In that paper we used special coordinates introduced by Ehlers and Kundt \cite{EK}. This reduction involved second-class constraints and complicated Dirac bracket relations.  Although the formulation is not a comfortable point of departure for loop quantization the Dirac brackets give analogs of the commutation relation of a linear self-dual field \cite{JF} with gravitational corrections that suggest corrections to $\hbar$ or the gravitational constant rather than a modified wave speed \cite{we}. In the present approach we formulate the reduction of the Gowdy model to pp-waves by means of a set of first-class constraints, so that the system becomes accessible to loop quantum techniques. A direct Hamiltonian formulation of pp-waves can be found in \cite{Bal}.

In this paper we review the formalism of Ref. \cite{BD1} in the next section.  Then we show in Section \ref{nullkilling} that the existence of a null Killing vector introduces constraints on the canonical variables. The new set of constraints for the model are shown to be first class in Section \ref{constraints}.  Finally, initial steps toward quantization are carried out in Section \ref{quantization}.
 
\section{The polarized Gowdy model in Ashtekar-Barbero variables}

The vacuum Gowdy models are characterized by closed spatial topologies and two, spatial commuting Killing vectors.  When the Killing vectors are orthogonal, then the model is ``polarized".  For our purposes it is convenient to employ the same formalism for pp-waves as used for polarized Gowdy models in Ashtekar-Barbero variables as formulated by Banerjee and Date  \cite{BD1}.  The model has to be adapted to the setting of pp-waves.  We replace the angular variable $\theta$ on $S^1$ of the Gowdy model by the variable $z$ on $\Re$. We assume homogeneity in the $x,y$ plane and wave propagation in the $z$ direction. We choose adapted spatial triads with one leg in the $z$ direction and two arbitrary orthogonal vectors in the homogeneous $x,y$ plane. Densitized inverse triad variables are denoted by ${E^a}_i$, $a=x,y$ and $i=1,2$. As in \cite{BD1} we write
\begin{equation}
{\cal E}:={E^z}_3
\end{equation}
and introduce ``polar coordinates" for the triad vectors in the $x,y$ plane,
\begin{equation}
\begin{array}{ll}
{E^x}_1=E^x\cos\eta, & {E^x}_2=E^x\sin\eta,\\
{E^y}_1=-E^y\sin\eta, & {E^y}_2=E^y\cos\eta.
\end{array}
\end{equation}
All variables depend only on $z$ and the time variable $t$. The gauge group $SU(2)$ of triad rotations has been reduced to $U(1)$ rotations in the $x,y$ plane, described by the dependence of the angle $\eta$ on $z$. In terms of these variables the spatial metric is
\begin{equation}\label{m}
{\rm d}s^2=h_{ab} {\rm d}x^a {\rm d}x^b = {\cal E}\,\frac{E^y}{E^x}\,{\rm d}x^2+{\cal
E}\,\frac{E^x}{E^y}\,{\rm d}y^2+\frac{E^xE^y}{\cal E}\,{\rm d}z^2.
\end{equation}

Like the polarized Gowdy model of Ref.\cite{BD1} the pp-wave space-times with either left of right moving waves are globally hyperbolic solutions to Einstein's equations that have two independent, orthogonal, commuting spatial Killing vectors.   Unlike this Gowdy model, however, the pp-wave space-times are characterized  by a null Killing field.   To apply loop quantization techniques it is useful to introduce a hypersurface-orthogonal time coordinate and ask, under what conditions does the model describe pp-wave space-times?  The answer will be in the form of a system of first class constraints in terms of the same phase space as the polarized Gowdy model.

In the polarized Gowdy model the extrinsic curvature turns out to be diagonal, its components are denoted by $K_x$, $K_y$, and $K_z$. The curvatures $K_x$ and $K_y$ turn out to be the corresponding diagonal elements of the Ashtekar-Barbero connection, divided by the Barbero-Immirzi parameter $\gamma$. As in Ref. \cite{BD1}, it is convenient to work with the 8-dimensional phase space $\left\{ (K_x,E^x), (K_y,E^y), ({\cal E},{\cal A}), (\eta,P) \right\}$, where $\cal A$ is the component ${A_z}^3$ of the connection, divided by $\gamma$, and the momentum $P$ conjugate to $\eta$ is constructed from the connection. The quantities $E^a$, $\cal A$ and $P$ transform as scalar densities, while the quantities $K_a$, $\cal E$ and $\eta$ as scalars under diffeomorphisms along the $z$ axis. The fundamental Poisson brackets are 
\begin{equation}
\{K_a(z),E^b(z')\}=\kappa\delta_a^b\delta(z-z'), \hspace{2mm}
\{{\cal A}(z),{\cal E}(z')\}=\kappa\delta(z-z'), \hspace{2mm}
\{\eta(z),P(z')\}=\kappa\gamma\delta(z-z')
\end{equation}
where $\kappa=8\pi G_{\rm Newton}$ is the gravitational constant.

In terms of these variables we have the following set of first-class constraints for GR: The Gau\ss\ constraint
\begin{equation}
G=\frac{1}{\kappa\gamma}\left({\cal E}'+P\right),
\end{equation}
(the prime denotes the derivative with respect to $z$), which generates rotations in the $x,y$ plane, the diffeomorphism constraint
\begin{equation}
C=\frac{1}{\kappa}\left[K_x'E^x+K_y'E^y-{\cal E}'{\cal
A}+\frac{\eta'}{\gamma}\,P\right],
\end{equation}
generating diffeomorphisms along the $z$-axis, and the Hamiltonian
constraint
\begin{equation}\label{H}
\begin{array}{ll}
H\!\!&=-\displaystyle\frac{1}{\kappa\sqrt{{\cal
E}E^xE^y}}\left[E^xK_xE^yK_y+(E^xK_x+E^yK_y){\cal E}\left({\cal
A}+\frac{\eta'}{\gamma}\right)
-\displaystyle\frac{1}{4}{{\cal E}'}^2-{\cal EE}''\rule{0mm}{8mm}\right.\\[5mm]
&-\left.\displaystyle\frac{1}{4}{\cal
E}^2\left[\left(\ln\displaystyle\frac{E^y}{E^x}\right)'\right]^2
+\displaystyle\frac{1}{2}{\cal EE}'(\ln
E^xE^y)'\right]-\displaystyle\frac{\kappa}{4\sqrt{{\cal
E}E^xE^y}}\,G^2-\gamma\left(\sqrt{\displaystyle\frac{\cal
E}{E^xE^y}}\,G\right)'.\end{array}
\end{equation}

In the following calculations it is convenient to use weakly equivalent forms of the last two constraints. Using the Gau\ss\ constraint the diffeomorphism constraint can be rewritten in the form
\begin{equation}
C=\frac{1}{\kappa}\left[K_x'E^x+K_y'E^y-{\cal E}'\left({\cal
A}+\frac{\eta'}{\gamma}\right)\right]+\eta'G.
\end{equation}
We make use of the weakly equivalent constraint
\begin{equation}
\bar C :=K_x'E^x+K_y'E^y-{\cal E}'\left({\cal
A}+\frac{\eta'}{\gamma}\right) \approx \kappa C.
\end{equation}
This alternate form of the diffeomorphism constraint, $\bar{C}$, is proportional to $C$ modulo $G$. Similarly we define a form of the Hamiltonian constraint $\bar H$,
\begin{equation}
{\bar H} := - E^xK_xE^yK_y- (E^xK_x+E^yK_y){\cal E}\left({\cal A}+\frac{\eta'}{\gamma}\right) + \frac{1}{4}{{\cal E}'}^2 + {\cal EE}'' + \frac{1}{4}\left[ {\cal E} \left(\ln \frac{E^y}{E^x}\right)'\right]^2 -\frac{1}{2}{\cal EE}'(\ln E^xE^y)'
\end{equation}
weakly equal to $\kappa\sqrt{{\cal E}E^xE^y} H$.

\section{Reduction to pp-waves}
\label{nullkilling}

We want to consider finite pulses of plane-fronted, parallel waves, traveling either in the positive or in the negative $z$-direction through flat space. Such waves are characterized by a null Killing vector field in the direction of propagation. In order to formulate the null Killing vector field we add an orthogonal time coordinate to the spatial manifold with the metric (\ref{m}). Using a lapse function $N=N(t,z)$ we have
\begin{equation}
\label{4m}
{\rm d}s^2 = - N^2 {\rm d}t^2 + {\cal E}\,\frac{E^y}{E^x}\,{\rm d}x^2+{\cal
E}\,\frac{E^x}{E^y}\,{\rm d}y^2+\frac{E^xE^y}{\cal E}\,{\rm d}z^2 .
\end{equation}
The existence of the null Killing field satisfying $\nabla_{(\mu} k_{\nu)}=0$ gives rise to constraints on the phase space variables.  Using the relations worked out in Appendix \ref{appenA} we find two new constraints
\begin{eqnarray}
&&U_x=E^xK_x-\frac{1}{2}\,{\cal E}'-\frac{1}{2}\,{\cal
E}\left(\frac{{E^y}'}{E^y}-\frac{{E^x}'}{E^x}\right)=0,\label{Ux}\\
&&U_y=E^yK_y-\frac{1}{2}\,{\cal E}'+\frac{1}{2}\,{\cal
E}\left(\frac{{E^y}'}{E^y}-\frac{{E^x}'}{E^x}\right)=0.\label{Uy}
\end{eqnarray}
where the minus sign in $k^{\mu}$ in (\ref{kill}) was chosen \footnote{One could of course choose the other sign to obtain right-moving waves.  Since we will only be including waves moving in one direction we can neglect these constraints.}.  The two relations render the diffeomorphism constraint and the Hamiltonian constraint equivalent; rewriting $\bar C$ and $\bar H$ in terms of $U_x$ and $U_y$ gives
\begin{equation}\label{C}
{\bar C} \approx - \frac{1}{\cal E}\:\bar H={\cal E}''+\frac{\cal
E}{2}\left[\left(\ln\frac{E^y}{E^x}\right)'\right]^2-\frac{{\cal
E}'}{2}(\ln E^xE^y)'-{\cal E}'\left({\cal
A}+\frac{\eta'}{\gamma}\right).
\end{equation}
The weak equality in (\ref{C}) implies that modulo $U_x$, $U_y$, and the Gau\ss\ constraint
\[
C \approx -\sqrt{\frac{E^xE^y}{\cal E}}\,H=-\sqrt{-g_{zz}}\,H.
\]
This means that if we choose the lapse function $N = \sqrt{\frac{E^xE^y}{\cal E}}$ then the time evolution generated by $H$ is equivalent to a spatial diffeomorphism generated by $C$, i.\,e. time derivatives are equal to minus $z$-derivatives and the variables depend only on $t-z$. The waves travel without dispersion. 

In coordinates, such that $g_{zz}=1$, i.\,e. ${\cal E}=E^xE^y$ and ${\cal A}=\eta=0$, the constraint equation $\bar C=0$ reduces to ${E^x}''E^y+E^x{E^y}''=0$ \cite{we}. In the formulation by Ehlers and Kundt \cite{EK} (see also \cite{MTW}), where $E^x=Le^{-\beta}$ and $E^y=Le^\beta$, this becomes the the Einstein equation for pp-waves
\begin{equation}
L''+L(\beta')^2=0,
\end{equation}
where the prime and the functional dependence are in terms of $t \pm z$.

\section{The Constraint Algebra}
\label{constraints}

We now investigate the possibility of imposing the pp-wave conditions (\ref{Ux}) and (\ref{Uy}) as constraints, augmenting the GR constraints $G$, $C$, and $H$. It turns out that the whole set of constraints is reducible and that the resulting reduced system is first class. 

The expressions $U_x$ and $U_y$, smeared out with test functions, are
\begin{equation}
U_a[f]:=\int{\rm d}z\,f(z)U_a(z).
\end{equation}
The Poisson bracket structure may be summarized in the following matrix
\begin{equation}
\left(\begin{array}{cc} \{U_x[f],U_x[g]\} & \{U_x[f],U_y[g]\} \\
\{U_y[f],U_x[g]\} & \{U_y[f],U_y[g]\}
\end{array}\right)=\frac{1}{2}\left(\begin{array}{rr} 1 & -1 \\ -1 &
1\end{array}\right)\int{\rm d}z\,(f'g-fg')\,{\cal E}.
\end{equation}
This matrix is diagonalized by the linear combinations
\begin{equation}
U_+ := U_x+U_y, \hspace{1cm} U_- := U_x-U_y,
\end{equation}
explicitly
\begin{eqnarray} 
U_+ &=& E^xK_x+E^yK_y-{\cal E}' \text{  and} \label{Updef} \\
U_- &=& E^xK_x-E^yK_y-{\cal E}\left(\ln\frac{E^y}{E^x}\right)'. \label{Umdef}
\end{eqnarray}
The algebra of these constraints is
\begin{eqnarray}
\label{PB}
\{U_+[f],U_+[g]\} &=& \{U_+[f],U_-[g]\}=0, \text{  and} \\
\{U_-[f],U_-[g]\} &=& 2\!\int\!\!{\rm d}z\,(f'g-fg')\,{\cal E}.
\end{eqnarray}
In local form, the non-vanishing Poisson brackets are
\begin{equation}\label{U-}
\{U_-(z),U_-(z')\}=-2\,[{\cal E}(z)+{\cal E}(z')]\,\delta'(z-z'),
\end{equation}
accordingly the constraints $U_-(z)$ are second-class. But the full set of constraints $G,C,H, U_+,$ and  $U_-$ are reducible. In Appendix B we derive the relation
\begin{equation}\label{U}
U_-^2=U_+^2+2{\cal E}\left[\left(\ln\displaystyle\frac{\cal
E}{E^xE^y}\right)'+2\left({\cal
A}+\displaystyle\frac{\eta'}{\gamma}\right)\right]U_++4{\cal
E}U_+'+4\bar H-4{\cal E}{\bar C}
\end{equation}
showing that $U_-$ is identically zero on the $\{C=0, H=0, U_+=0 \}$-constraint surface. The constraint $U_-$ depends nonlinearly on these constraints.

The constraint $U_-$ generates no further secondary constraints since the Poisson brackets of $U_-$ with $G$, $C$
and $H$ weakly or strongly vanish,
\begin{equation}\begin{array}{l}
\{U_-[f],G[g]\}=0, \hspace{8mm}
\{U_-[f],C[g]\}=-\displaystyle\frac{1}{\kappa}\,U_-[f'g] \approx 0 ,\\[5mm]
\{U_-[f],H[g]\}=\displaystyle\frac{1}{\kappa}\,U_-\left[\sqrt{\frac{\cal
E}{E^xE^y}}\,f'g+\frac{{\cal E}'fg}{\sqrt{{\cal
E}E^xE^y}}\right]-\frac{1}{\kappa}\,U_+\left[\sqrt{\frac{\cal
E}{E^xE^y}}\,fg\right] \approx 0;
\end{array}
\end{equation}
$U_-$ is gauge invariant and conserved under spatial diffeomorphisms and time evolution.

Remarkably the constraint $U_+$ weakly Poisson-commutes with $G$, $C$, and $H$,
\begin{eqnarray}
&&\{U_+[f],G[g]\}=0,\\
&&\{U_+[f],C[g]\}=-\frac{1}{\kappa}\,U_+[f'g]\approx0,\\
&&\{U_+[f],H[g]\}=\frac{1}{\kappa}\,U_+\left[\sqrt{\frac{\cal
E}{E^xE^y}}\,f'g\right]-H[fg]\approx0.
\end{eqnarray}
Thus, $U_+$ can be added as another first-class constraint. The model now has the enlarged Poisson bracket algebra of the first-class constraints $G$, $C$, $H$, and $U_+$ that includes the standard algebra of constraints, adapted to the Gowdy model \cite{BD1},
\begin{equation}
\begin{array}{c}
\{G[f],G[g]\}=\{G[f],H[g]\}=0,\hspace{5mm}\{G[f],C[g]\}=-G[f'g],\\[4mm]
\{C[f],C[g]\}=C[fg'-f'g],\hspace{5mm}
\{C[f],H[g]\}=H[fg'],\\[4mm]
\{H[f],H[g]\}=C\left[(fg'-f'g)\displaystyle\frac{\cal
E}{E^xE^y}\right],
\end{array}
\end{equation}
This sets the stage for quantization as now the algebra is first class, unlike in the analysis of Ref. \cite{we}.

The number of four first-class constraints is the maximum that can be imposed on a system with four canonical degrees of freedom. According to their nature as generators of gauge transformations the three sets of standard constraints of GR reduce the number of canonical degrees of freedom to one (two phase space functions). The constraints $U_+(z)$, on the other hand, reduce the two phase space degrees of freedom at every point - one field variable and its conjugate momentum - to one. By doing this $U_+$ reduces the physical degrees of freedom and but does not fall into the class of gauge generators, even though it is first-class, thus providing another counterexample to the Dirac conjecture.

The quantity $E^xK_x+E^yK_y$ appearing in $U_+$, is the densitized trace of the extrinsic curvature of the wave fronts, when embedded into three-dimensional space-time. It has a geometric interpretation as the expansion or contraction of an transverse area element under time evolution. Likewise, the difference $E^xK_x-E^yK_y$ in the constraint $U_-$ has an interpretation as shear of transverse area elements. The analysis of the this section shows that area expansion can be integrated into the first-class constraint system, whereas shear cannot. 

\section{Preparation for quantization}
\label{quantization}

For quantization it is important that $U_+$ can be given a meaning as a well-defined operator. Indeed all the constraints can be formulated in a way that anticipates the construction of the corresponding loop quantum operators. Both $E^xK_x+E^yK_y$ and ${\cal E}'$ are scalar densities, which can  be naturally integrated along $z$, so to construct an operator we have to integrate them. The integral over some interval $\cal I$ of the coordinate $z$ is
\begin{equation}\label{K}
U_+[{\cal I}]=\int_{\cal I}\!{\rm d}z(E^xK_x+E^yK_y)-{\cal
E}_++{\cal E}_-,
\end{equation}
where ${\cal E}_\pm$ are the values of $\cal E$ at the endpoints of $\cal I$. $\cal E$ has a meaningful operator equivalent in the adapted LQG framework of Ref. \cite{BD2}. In analogy to full LQG the integral in (\ref{K}) can be obtained as the Poisson bracket
\begin{equation}
\left\{\int_{\cal I}{\rm d}z \frac{E^xK_xE^yK_y}{\sqrt{{\cal
E}E^xE^y}},\int_{\cal I} {\rm d}z' \,\sqrt{{\cal
E}E^xE^y}\right\}=\frac{1}{2}\int_{\cal I} {\rm d}z(E^xK_x+E^yK_y).
\end{equation}
The first expression in the Poisson bracket is the first term of the kinetic term in the Hamiltonian constraint of (28) in Ref. \cite{BD2}. The second expression is the volume of a sandwich of space, constructed from a fiducial (unit) area in the $x,y$ plane as basis and an interval $\cal I$ in the $z$-direction. Both expressions have operator equivalents in standard LQG, for the present case we find the corresponding operators in \cite{BD2}, equations (31) and (32). Now we are in a position to express all first-class constraints in terms of loop quantum operators, acting on one-dimensional spin network states, as demonstrated in \cite{BD2}.

\section{Conclusion and outlook}
In this paper we showed how the polarized Gowdy model can be further reduced to describe pp-waves.  Somewhat surprisingly, the constraints from GR augmented by those derived from the Killing equations are reducible and that the reduced constraints form a first class system, which is amenable to loop quantum gravity quantization techniques.  Much of the preparation of the quantization is carried out in Ref. \cite{BD2}.  Work is underway on quantization of the model, which if all goes well, will provide a framework in which to perform {\em ab initio} investigations on Lorentz violation and deformation for pp-wave space-times.

In classical GR the existence of a null Killing vector field in the direction of propagation of gravitational pp-waves guarantees dispersion-free propagation of such waves at a constant speed. In the present paper we show that a linear combination of two Killing equations is implemented as a first class constraint, which  describes the expansion of null geodesics. The full information of the Killing equations, including shear, namely $U_+$ and $U_-$, that would render the Hamiltonian and the diffeomorphism constraint weakly equivalent, cannot be imposed in the form of first-class constraints. More specifically, if we exchange the first-class set $G$, $C$, $H$, $U_+$ for $G$,
$C$, $U_+$ and $U_-$, to make $H$ equivalent to $C$ (see (\ref{C})), the system becomes second-class. So, in order to be able to apply loop quantization techniques, we keep both $C$ and $H$ as independent constraints and the classical equivalence of time derivatives to space derivatives is not manifest in the constraint algebra. 

If we insist on implementing the full contents of the Killing equations as constraints, we are led to handle the second-class constraints $U_-$ with the method of Dirac brackets and impose them strongly. This program was carried out in detail in \cite{we} in a non-diffeomorphism invariant manner.  

An ongoing quantum investigation in the spirit of Ref. \cite{BD2} will bring more clarity about possible dispersion of pp-waves in LQG. The results will be interesting to compare with Ref. \cite{MB1}, which deals with gravitational wave dispersion in a cosmological background using perturbative techniques. In isolating pure gravitational wave propagation in one direction we have ended up with a more simple model. With the higher degree of simplification we may expect complementary results to Ref. \cite{MB1}.

\appendix

\section{Space-time metric, Killing equation} 
\label{appenA}

In order to formulate a null Killing vector field we have to construct a space-time manifold. To this purpose we add an orthogonal time coordinate to the space-like manifold with the metric (\ref{m}). The latter one becomes supplemented by a lapse function $N=N(t,z)$, so that
\begin{equation}
{\rm d}s^2 = - N^2 {\rm d}t^2 + {\cal E}\,\frac{E^y}{E^x}\,{\rm d}x^2+{\cal
E}\,\frac{E^x}{E^y}\,{\rm d}y^2+\frac{E^xE^y}{\cal E}\,{\rm d}z^2 .
\end{equation}
The non-vanishing components of the Levi-Civit\`{a} connection are the following Christoffel symbols
\[\begin{array}{lll}
\Gamma^0_{00}=\displaystyle\frac{\dot N}{N}, &
\Gamma^3_{00}=\displaystyle\frac{{\cal E}NN'}{E^xE^y}, &
\Gamma^0_{03}=\Gamma^0_{30}=\displaystyle\frac{N'}{N},\\[4mm]
\Gamma^0_{33}=\displaystyle\frac{1}{2N^2}  \partial_t \left(\frac{E^xE^y}{\cal E}\right), & 
\Gamma^3_{33}=\displaystyle\frac{1}{2}\frac{\cal E}{E^xE^y}\left(\displaystyle\frac{E^xE^y}{\cal E}\right)', &
\Gamma^3_{03}=\Gamma^3_{30}=\displaystyle\frac{1}{2}\frac{\cal E}{E^xE^y} \partial_t \left(\displaystyle\frac{E^xE^y}{\cal E}\right),
\end{array}
\]
\[
\begin{array}{ll}
\Gamma^0_{11}=\displaystyle\frac{1}{2N^2} \partial_t \left({\cal E}\frac{E^y}{E^x}\right), &
\Gamma^1_{01}=\Gamma^1_{10}=\displaystyle\frac{1}{2}\frac{E^x}{{\cal E}E^y} \partial_t \left({\cal E}\frac{E^y}{E^x}\right),\\[4mm]
\Gamma^3_{11}=-\displaystyle\frac{1}{2}\frac{\cal E}{E^xE^y}\left({\cal E}\frac{E^y}{E^x}\right)', &
\Gamma^1_{13}=\Gamma^1_{31}=\displaystyle\frac{1}{2}\frac{E^x}{{\cal E}E^y}\left({\cal E}\frac{E^y}{E^x}\right)',\\[4mm]
\Gamma^0_{22}=\displaystyle\frac{1}{2N^2} \partial_t \left({\cal E}\frac{E^x}{E^y}\right) , &
\Gamma^2_{02}=\Gamma^2_{20}=\displaystyle\frac{1}{2}\frac{E^y}{{\cal E}E^x} \partial_t \left({\cal E}\frac{E^x}{E^y}\right),\\[4mm]
\Gamma^3_{22}=-\displaystyle\frac{1}{2}\frac{\cal E}{E^xE^y}\left({\cal E}\frac{E^x}{E^y}\right)', &
\Gamma^2_{23}=\Gamma^2_{32}=\displaystyle\frac{1}{2}\frac{E^y}{{\cal E}E^x}\left({\cal E}\frac{E^x}{E^y}\right)'.
\end{array}
\]
With the above four-metric a null vector field in the $z$-direction is of the form
\begin{equation}\label{kill}
k^\mu=\left( \sqrt{\frac{E^xE^y}{\cal E}}\,k,0,0,\pm Nk\right),
\end{equation}
with $k$ being a function of $t$ and $z$. For $k^\mu$ to be a Killing vector field, the Killing equations
$\nabla_{(\mu} k_{\nu)} =0$ in terms of covariant derivatives with respect to the Levi-Civit\`{a} connection must hold. From $\nabla_t k^t =0$ we obtain the time derivative of $k$,
\begin{equation}
\dot k=-\left[\frac{1}{2}  \frac{\cal E}{E^xE^y} \partial_t \left(\frac{E^xE^y}{\cal E}\right) \pm \sqrt{\frac{\cal
E}{E^xE^y}}\,N'\pm\frac{\dot N}{N}\right]k,
\end{equation}
whereas from $\nabla_z k^z =0$ we get the $z$-derivative,
\begin{equation}
k'=-\left[\frac{1}{2N}\sqrt{\frac{\cal
E}{E^xE^y}} \partial_t \left(\frac{E^xE^y}{\cal
E}\right)+\frac{1}{2}\frac{\cal E}{E^xE^y}\left(\frac{E^xE^y}{\cal
E}\right)'+\frac{N'}{N}\right]k.
\end{equation}
With these derivatives the equation $\nabla_{(t} k_{z)}=0$ reduces to
an identity.

Before imposing the remaining two non-trivial Killing equations $\nabla_a k_a=0$ we introduce the extrinsic curvature. In the case of a vanishing shift vector it is
\begin{equation}
K_{ab}=-\frac{1}{2N}\,\dot{g_{ab}},
\end{equation}
so that we can express all time derivatives in the Christoffel symbols in terms of it. In the case of the polarized Gowdy model \cite{BD1} the diagonal components of the Ashtekar-Barbero connection, which
are canonically conjugate to $E^x$ and $E^y$, are - modulo the
Barbero-Immirzi parameter - equal to the extrinsic curvature
components.

In the following we will use the components ${K_a}^i={e_b}^iK^b_a$,
converted with the un-densitized radial triad components
\begin{equation}
\begin{array}{l}
{e_x}^1=\sqrt{{\cal
E}\displaystyle\frac{E^y}{E^x}}\,\cos\displaystyle\frac{\eta}{2},\hspace{5mm}{e_x}^2=
\sqrt{{\cal E}\displaystyle\frac{E^y}{E^x}}\,\sin\displaystyle\frac{\eta}{2}\\[5mm]
{e_y}^1=-\sqrt{{\cal
E}\displaystyle\frac{E^x}{E^y}}\,\sin\displaystyle\frac{\eta}{2},\hspace{5mm}
{e_y}^2=\sqrt{{\cal
E}\frac{E^x}{E^y}}\,\cos\displaystyle\frac{\eta}{2},\hspace{5mm}
{e_z}^3=\sqrt{\displaystyle\frac{E^xE^y}{\cal E}},
\end{array}
\end{equation}
corresponding to $E^x$, $E^y$, and $\cal E$, respectively. In terms
of these components we have
\begin{eqnarray}
K_x &:=&\sqrt{({K_x}^1)^2+({K_x}^2)^2}=\frac{1}{2N} \sqrt{ \frac{ E^x}{{\cal E} E^y} } \partial_t \left({\cal
E}\frac{E^y}{E^x}\right)
\\[3mm]
K_y &:=& \sqrt{({K_y}^1)^2+({K_y}^2)^2}=\frac{1}{2N} \sqrt{\frac{E^y}{{\cal E} E^x}} \partial_t \left({\cal E}\frac{E^x}{E^y}\right)
\\[3mm]
{K_z}^3 &=& \frac{1}{2N}\frac{\cal E}{E^xE^y} \partial_t \left(\frac{E^xE^y}{\cal
E}\right)=\frac{1}{2N}\left(\frac{\dot{E^x}}{E^x}+\frac{\dot{E^y}}{E^y}-
\frac{\dot{\cal E}}{\cal E}\right).
\end{eqnarray}
To express $\nabla_a k_{a}=0$ in terms of canonical variables we can use $\Gamma^0_{11}$ and $\Gamma^0_{22}$ in the form
\begin{equation}
\Gamma^0_{11}=\frac{K_x}{N} \sqrt{ \frac{{\cal E} E^y}{E^x}} \text{  and  } \Gamma^0_{22}=\frac{K_y}{N}\sqrt{ \frac{{\cal E} E^x}{E^y} }.
\end{equation}
The Killing equations are independent of $k$
\begin{eqnarray}
\mp E^xK_x &=& \frac{1}{2} {\cal E} \left( \frac{{\cal E}'}{{\cal E}} + \frac{{E^y}'}{E^y} - \frac{{E^x}'}{E^x}\right)
\\
\mp E^yK_y &=& \frac{1}{2} {\cal E} \left( \frac{{\cal E}'}{{\cal E}} - \frac{{E^y}'}{E^y}+ \frac{{E^x}'}{E^x} \right),
\end{eqnarray}
which are the constraints (\ref{Ux}) and (\ref{Uy}).

\section{The relation between $U_-$, $U_+$, $\bar C$ and $ \bar H$}

In this appendix we show some steps of the straightforward demonstration of the dependence of $U_-$, $U_+$, $\bar C$ and $ \bar H$.  From the expressions for $U_-$ and $U_+$ in (\ref{Updef}) and (\ref{Umdef}) we see that  
\begin{equation}\label{45}
E^xK_x=\frac{1}{2}\left[U_++U_-+{\cal E}'+{\cal
E}\left(\ln\frac{E^y}{E^x}\right)'\right]
\end{equation}
and
\begin{equation}\label{46}
E^yK_y=\frac{1}{2}\left[U_+-U_-+{\cal E}'-{\cal
E}\left(\ln\frac{E^y}{E^x}\right)'\right]
\end{equation}
give
\begin{equation}
E^xK_x+E^yK_y=U_++{\cal E}'
\end{equation}
and
\begin{equation}
\begin{array}{ll}E^xK_xE^yK_y\!\!&=\displaystyle\frac{1}{4}\,U_+^2+\frac{1}{2}\,{\cal
E}'U_+-\frac{1}{4}\,U_-^2-\frac{1}{2}\,{\cal
E}\left(\ln\frac{E^y}{E^x}\right)'U_-\\[4mm]
&+\displaystyle\frac{1}{4}\,({\cal E}')^2-\frac{1}{4}\,{\cal
E}^2\left[\left(\ln\displaystyle\frac{E^y}{E^x}\right)'\right]^2.
\end{array}
\end{equation}
Using these results in $\bar H$ produces
\begin{equation}\label{hbar}
\begin{array}{ll}
\bar H\!\!&=-\displaystyle\frac{1}{4}\,U_+^2-\frac{1}{2}\,{\cal E}'U_++\frac{1}{4}\,U_-^2+\frac{1}{2}\,{\cal
E}\left(\ln\frac{E^y}{E^x}\right)'U_--{\cal E}\left({\cal A}+\displaystyle\frac{\eta'}{\gamma}\right)U_+  \\[5mm]
&+\,{\cal E\,E}''+\displaystyle\frac{1}{2}\,{\cal E}^2\left[\left(\ln\displaystyle\frac{E^y}{E^x}\right)'\right]^2-\frac{1}{2}\,
{\cal E\,E}'(\ln E^xE^y)'-{\cal E\,E}'\left({\cal A}+\displaystyle\frac{\eta'}{\gamma}\right).
\end{array}\end{equation}
Differentiating $U_+$ and inserting into $\bar C$ gives
\begin{equation}
\begin{array}{ll}
\bar C\!&=U_+'-\displaystyle\frac{1}{2}\,(\ln
E^xE^y)'U_++\frac{1}{2}\left(\ln\displaystyle\frac{E^y}{E^x}\right)'U_-\\[4mm]
&-\displaystyle\frac{1}{2}\,{\cal E}'(\ln E^xE^y)'+{\cal
E}''+\frac{1}{2} {\cal E} \left[\left(\ln\displaystyle\frac{E^y}{E^x}\right)'\right]^2-{\cal
E}'\left({\cal A}+\displaystyle\frac{\eta'}{\gamma}\right).
\end{array}\end{equation}
From this we can eliminate ${\cal E} {\cal E}'\left({\cal A}+\frac{\eta'}{\gamma}\right)$ from $\bar H$. This substitution is sufficient to express $U_-$ completely in terms of the first-class constraints $\bar C$, $\bar
H$, and $U_+$, resulting in relation (\ref{U}). Thus, we have a quadratic equation for $U_-$, the solutions of which are of course also zero modulo the first-class constraints.

\begin{acknowledgements} The work was supported by the Ministry of Education of the Czech Republic, contract no. MSM 0021622409.
\end{acknowledgements}

\end{document}